\documentclass{svmult}

\usepackage{epsfig}
\usepackage{amsmath}
\usepackage{natbib}

\def\Rreal{{\rm I\kern-.2em R}}

\font\tenbi=cmmib10
\newfam\bifam \def\bi{\fam\bifam\tenbi} \textfont\bifam=\tenbi

\def\vecV{{\textstyle{\bi V}}}

\newcommand{\aap}{    {\rm Astron. Astrophys.}\ }

\newcommand{\apj}{    {\rm Astrophys. J.}\ }
\newcommand{\apjl}{   {\rm Astrophys. J. Lett.}\ }

\newcommand{\solphys}{{\rm Solar Phys.}\ }

\newcommand{\ssr}{    {\rm Space Sci. Rev.}\ }

\usepackage{amsfonts}   
\begin{document}
\title*{Local Helioseismology of Emerging Active Regions: A Case Study}
\toctitle{Local Helioseismology of Emerging Active Regions: A Case Study}
%
%
\titlerunning{Local Helioseismology of Emerging Active Regions}
%

\author{Alexander G. Kosovichev,  Junwei Zhao, \& Stathis Ilonidis}
\authorrunning{Alexander G. Kosovichev, Junwei Zhao, \& Stathis Ilonidis}

\institute{New Jersey Institute of Technology, Newark, NJ 07103, USA;\\Stanford University, Stanford, CA 94305, USA\\E-mail: \texttt{alexander.g.kosovichev@njit.edu}}


\maketitle              

\begin{abstract}
\noindent 
Local helioseismology provides a unique opportunity to investigate the subsurface structure and dynamics of active regions and their effect on the large-scale flows and global circulation of the Sun. We use measurements of plasma flows in the upper convection zone, provided by the Time-Distance Helioseismology Pipeline developed for analysis of solar oscillation data obtained by the Helioseismic and Magnetic Imager (HMI) on board of Solar Dynamics Observatory (SDO), to investigate the subsurface dynamics of emerging active region NOAA 11726.  The active region emergence was detected in deep layers of the convection zone about 12 hours  before the first bipolar magnetic structure appeared on the surface, and 2 days before the emergence of most of the magnetic flux.  The speed of emergence determined by tracking the flow divergence with depth is about $1.4$~km/s, very close to the emergence speed in the deep layers.  As the emerging magnetic flux becomes concentrated in sunspots local converging flows are observed beneath the forming sunspots. These flows are most prominent in the depth range 1-3~Mm, and remain converging after the formation process is completed. On the larger scale converging flows around active region appear as a diversion of the zonal shearing flows towards the active region, accompanied by formation of a large-scale vortex structure. This process occurs when a substantial amount of the magnetic flux emerged on the surface, and the converging flow pattern remains stable during the following evolution of the active region. The Carrington synoptic flow maps show that the large-scale subsurface inflows are typical for other active regions. In the deeper layers (10-13~Mm) the flows become diverging, and surprisingly strong beneath some active regions. In addition to the flows around active regions, the synoptic maps reveal a complex evolving pattern of large-scale flows on the scale much larger than supergranulation. 
\end{abstract}

\section{Introduction}\label{sec1}
Emergence and formation of magnetic active regions on the surface of the Sun is one of the central problems of solar physics. It is of the fundamental importance in astrophysics because active regions are one of the primary manifestations of the solar and stellar magnetism.  In addition, solar active regions (AR) are the major drivers of the solar variability, geospace and planetary space environments and space weather. Understanding of the emergence and evolution of active regions is a key to developing the knowledge and capability to detect and predict extreme conditions in space. The uninterrupted helioseismology and magnetic data from Solar Dynamics Observatory provide unique opportunities for comprehensive  studies that can uncover the basic mechanisms of active region formation, evolution, and their flaring and CME activity \citep{Scherrer2012}. 

Recent studies revealed that the emergence and magnetic structure of active regions are closely linked to the plasma flows on the surface and in the subsurface layers \citep{Hindman2009,Komm2011,Komm2012,Kosovichev2012,Kosovichev2006,Birch2013}. For example, diverging subsurface flows have been detected prior the emergence of active regions, shearing and twisting flows are found to be associated with flaring activity, large-scale converging flows formed around active regions affect the meridional circulation and along with the tilt of active regions (Joy’s law) are believed to be among primary factors determining the strength and duration of the future solar cycles. 

The conventional wisdom is that the active regions are a result of emergence of toroidal magnetic  flux ropes formed near the bottom of the convection zone by a dynamo process. This theory can explain the Joy's law and the absence of active regions at high latitudes providing the initial magnetic field strength is about 60~kG \citep{DSilva1993}, which greatly exceeds the equipartition field strength and has not been reproduced by the dynamo theories. Current 3D MHD global-Sun models computed in the anelastic approximation have shown that the dynamo-generated field can be organized in the form of flux tubes but on much larger scale \citep{Brun2004,Fan2014,Guerrero2016}. These models indicates that the active regions and sunspots are probably formed in the near-surface layers, but the anelastic approximation becomes invalid close to the surface, where compressibility effects play significant role. With the currently available computational resources the realistic compressible radiative MHD simulations  are capable to model only relatively shallow near-surface. These simulations have revealed a process of spontaneous formation of compact pore-like structures from initially distributed magnetic fields, maintained by converging downdrafts, however, other mechanisms of magnetic self-organization may be also involved \citep{Kapyla2016,Kitiashvili2010,Masada2016}. 

The magnetic self-organization process  is probably a key to understand the formation of sunspots and active regions. It involves a complex interaction of turbulence with magnetic field, but in all cases the large-scale flow pattern includes compact regions of converging downdrafts around magnetic structures in shallow $\sim 5$~Mm deep regions. In the deeper layers the flows are mostly diverging. This flow pattern corresponds to the Parker's cluster model of sunspots. It has been observed by the time-distance helioseismology analysis of the SOHO/MDI and Hinode/SOT data \citep{Zhao2001,Zhao2009b,Zhao2003}. The data analysis also showed that in the decaying sunspots the flows become diverging.  Other helioseismology methods, such as the ring-diagram analysis and the helioseismic holography, provide the subsurface flow maps with a lower resolution than the time-distance helioseismology, and did not confirm the existence of the converging downdrafts beneath the sunspots. Instead, they inferred diverging flows of a larger scale around sunspots over the whole depth range probed by these techniques \citep{Hindman2009}. 

 The current helioseismology measurements in regions of strong magnetic field are subject to significant systematic errors due to the uncertainties in the Doppler-shift measurements, large variations of the sound speed causing non-linear wave effects, non-uniform distribution of acoustic sources and MHD wave transformation effects. These uncertainties mostly affect inferences of the sound-speed distribution beneath the sunspots  \citep[for a recent review see][]{Kosovichev2012}. The helioseismic inferences  of subsurface flows are based on measuring and inverting the travel-time differences for the waves traveling along the same path in the opposite directions. Such reciprocal signals are less sensitive to the systematic uncertainties. However, a complete testing and calibration of the flow inferences based on numerical simulations of wave propagation in sunspots has not been completed. 

In this work we mostly focus on flows of active regions that are formed beneath sunspots, and, in particular, on the flow patterns during the formation and evolution of a large emerging active region. The primary goal is to investigate the process of formation of the large-scale converging flows that affect the meridional circulation and magnetic flux transport. As a case study we consider a large emerging active region NOAA 11726.

\section{Time-Distance Helioseismology from SDO}\label{sec2}

\begin{figure}[t]
\begin{center}
  \includegraphics[width=\linewidth]{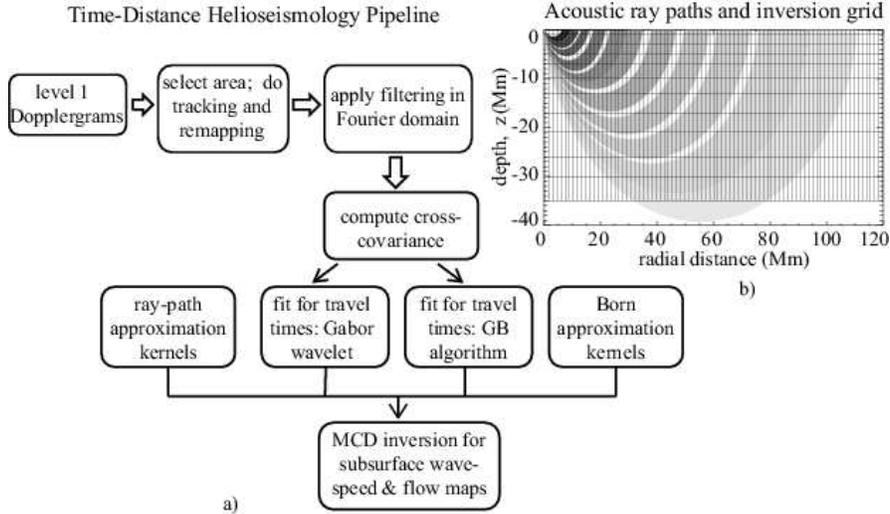}
  \caption{a) Diagram of the HMI Time-Distance Helioseismology pipeline implemented at the SDO Joint Science Operations Center \citep{Zhao2012}. b) a set acoustic paths and the inversion grid illustrating the measurement scheme used in the Time-Distance Pipeline. The inversion grid includes 11 depth intervals: 0-1, 1-3, 3-5, 5-7, 7-10, 10-13, 13-17, 17-21, 21-26, 26-30, and 30-35 Mm. The horizontal grid spacing is 0.12 degrees ($\sim 1.45$~Mm).}\label{fig1}
\end{center}
\end{figure}

The Helioseismic and Magnetic Imager (HMI) provides uninterrupted Dopplergrams with high spatial (0.5 arcsec per pixel, or 0.03 heliographic degrees at the disk center) and temporal (45 sec) resolutions. These data cover the whole spectrum of photospheric oscillations, and are ideal for local helioseismology studies. Developed as a part of the SDO helioseismology program, the Time-Distance Helioseismology Pipeline provides travel times of acoustic waves measured by two different methods, and also the maps of subsurface  flows and sound-speed perturbations obtained by using the Multi-Channel Deconvolution technique and two different types of sensitivity kernels derived from the ray-path and Born approximations (Fig.~\ref{fig1}a). Thus, the pipeline provides four different sets of inversions for the 3D flow velocities and wave-speed variations. Details of the pipeline procedures, and also the test results and estimation of errors are described by \citet{Couvidat2012} and \citet{Zhao2012}. The pipeline data have been used to determine the distributions of the flow vorticity and helicity, and also variations of the meridional circulation and zonal flows with the solar cycle \citep{Zhao2014,Kosovichev2016}. 

In addition, we have developed a complementary pipeline for tracking the subsurface dynamics of active regions. This pipeline takes the Carrington coordinates of active regions from the Solar Region Summary (SRS) database, compiled by the NOAA Space Weather Prediction Center (SWPC), and uses these coordinates as the central points of $30\times 30$-degree areas tracked for 10 days during their passage on the solar disk. This setup allows us to follow the evolution of active region areas even before the magnetic flux emergence and after the decay. The 3D subsurface flow maps are calculated from the tracked Dopplergrams that are remapped onto the heliographic coordinates using the Postel's projections. Each tracked, 8-hour long, datacube consists of 640 Dopplergrams of $512\times 512$ pixels with the spatial resolution of $0.06$ degree/pixel, and 45-sec time cadence. 

The tracked datacubes are  processed through the Time-Distance Helioseismology Pipeline (Fig.~\ref{fig1}a), and the output represents acoustic travel-time maps calculated  with $0.12$-deg sampling for the whole tracked areas ($256\times 256$ pixels). The travel-times are calculated for eleven annuli located at different distances from the central points  representing $2\times 2$-binned original Dopplergram pixels. The signals of acoustic waves traveling between the central points and the surrounding annuli  are calculated from the HMI Doppler velocity measurements as the corresponding cross-covariance functions. The cross-covariances are computed in the Fourier space, and  phase-space filters are applied to isolate the signals corresponding to each  of the travel distances. 

The travel times are calculated by two different methods: 1) the Gabor wavelet fitting \citep{Kosovichev1997a} and 2) a  cross-correlation with reference cross-covariance functions obtained by averaging over a large area \citep{Gizon2002}. Then, the travel times are used to infer the 3D maps of subsurface flows and sound-speed perturbations, by solving an inverse acoustic tomography problem. It is formulated in the form of linear integral equations the kernels of which are calculated by using the ray-path theory \citep{Kosovichev1997a} and the first Born approximation \citep{Birch2000,Birch2001a,Birch2001,Birch2004,Birch2007}. 

Regularized solutions to the inverse problem are determined by the Multi-Channel Deconvolution (MCD) method \citep{Jacobsen1999,Couvidat2006}, and the regularization parameters were chosen to suppress noise and represents a smooth solution. The depth coverage is illustrated in Fig.~\ref{fig1}b, which shows a vertical cut of the inversion grid together with the acoustic ray paths corresponding to the selected set of 11 annuli. Thus, the inversion results provide the 3D flow and sound-speed maps up to the depth of 30~Mm. However, for analysis we use only the top layers less than 20~Mm because the pipeline results become less robust as the `realization noise' of solar oscillations increases  with depth. 

The deeper interior of the Sun can be probed by increasing the spatial and temporal averaging of the oscillation cross-covariance function and extending the range of the acoustic ray paths. The deeper  penetrating waves travel to longer distances on the solar surface. The primary factors that restrict the resolving power of the time-distance helioseismology with depth are the increasing wavelength of acoustic waves and the increasing `realization noise'. The realization noise is a consequence of random excitation of solar acoustic waves \citep{Woodard1984}. Because the number of deeply-penetrating waves with long horizontal wavelength (that can be represented in terms of normal modes with relatively low angular degree) is smaller than the number of short acoustic waves  (high-degree modes) the realization noise increases for deep measurements. Nevertheless, by averaging the cross-covariance signals for two years \citep{Zhao2013} were able to measure the meridional flows up to the base of the convection zone.

\begin{figure}[t]
\begin{center}
  \includegraphics[width=\linewidth]{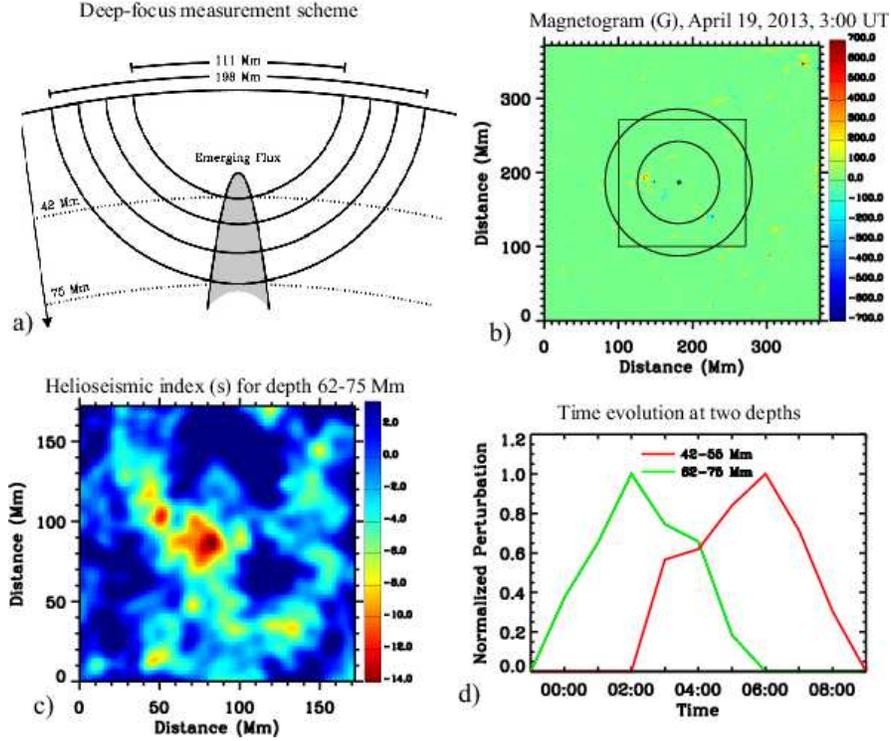}
  \caption{ Detection of emerging active region AR~11726: a) the HMI magnetogram of April 19, 2013, 3:00~UT, prior the emergence; the rings show the area where the acoustic oscillation signal was measured to detect the subsurface signal at the central point; the square is the area of the subsurface detection shown in panel $c$; b) schematic illustration of the deep-focus measurement scheme;  c) variations of a travel-time index showing the AR perturbation located the depth of $62-75$ Mm at April 19, 2013, 3:00~UT; d) variations of the helioseismic perturbation associated with the emerging AR at two different depths: $42-55$~Mm and $62-75$~Mm, as a function of time.   }\label{fig2}
\end{center}
\end{figure}

One of the great advantages of time-distance helioseismology is that observations of solar oscillations over the whole disk allow us to construct special measurement scheme to select and accumulate signals of acoustic waves traveling through particular regions below the surface. For instance, \citep{Ilonidis2013} developed a special deep-focusing procedure that is capable of detecting large emerging active regions 2 days before they emerge on the surface (Fig.~\ref{fig2}). This procedure will be illustrated in our case study of helioseismic diagnostics of emerging active region NOAA 11726. 

\section{Detection of active region before it becomes visible on the surface}\label{sec3}

The deep-focuse measurement scheme used for the subsurface detection of AR 11726 is shown in Fig.~\ref{fig2}a-b.  In this scheme the cross-covariance function is calculated for the ray paths with the lower turning points located at the depth $42-75$~Mm. This range of depth corresponds to the travel distance on the surface  of $111-198$~Mm. Thus, the cross-covariance function is calculated using the HMI Doppler velocity measurements located on the opposite side of the annuli shown in Fig.~\ref{fig2}b. In this case, the calculated cross-covariance function is mostly sensitive to perturbations located beneath the central point at the depth $42-75$~Mm, and is not affected by potential perturbations in the flux emergence area. The central point of the annuli is moved to different positions on the surface, and the  cross-covariance calculations are repeated. Phase shifts of the local cross-covariance function from the mean profile provide a map of subsurface perturbations at this depth. This approach was optimized for subsurface detection of emerging active regions by \citet{Ilonidis2013}.

Figure~\ref{fig2}c shows the distribution of an effective phase shift (`helioseismic index') in the range of depth $62-75$~Mm, measured on April 19, 2013, 03:00~UT when there was no significant magnetic flux on the surface (Fig.~\ref{fig2}b). Figure~\ref{fig2}d shows variations of the helioseismic perturbation associated with the emerging AR at two different depths: $62-75$~Mm and  $42-55$~Mm, as a function of time.  This approach allows us to track the development of subsurface perturbations with time, and estimate the speed of emergence, which in this case is about 1.4~km/s. The emergence of magnetic flux on the surface starts on the following day, and most of the magnetic flux emerged 2 days after it was first detected below the surface. Currently, this method provides an early detection only large active regions. Thus,  a strong helioseismic perturbation  below the surface observed prior the emergence may serve as a precursor of  large active regions on the solar surface.

\section{Subsurface dynamics of emerging active region}\label{sec4}

\begin{figure}
\begin{center}
\includegraphics[width=0.97\linewidth]{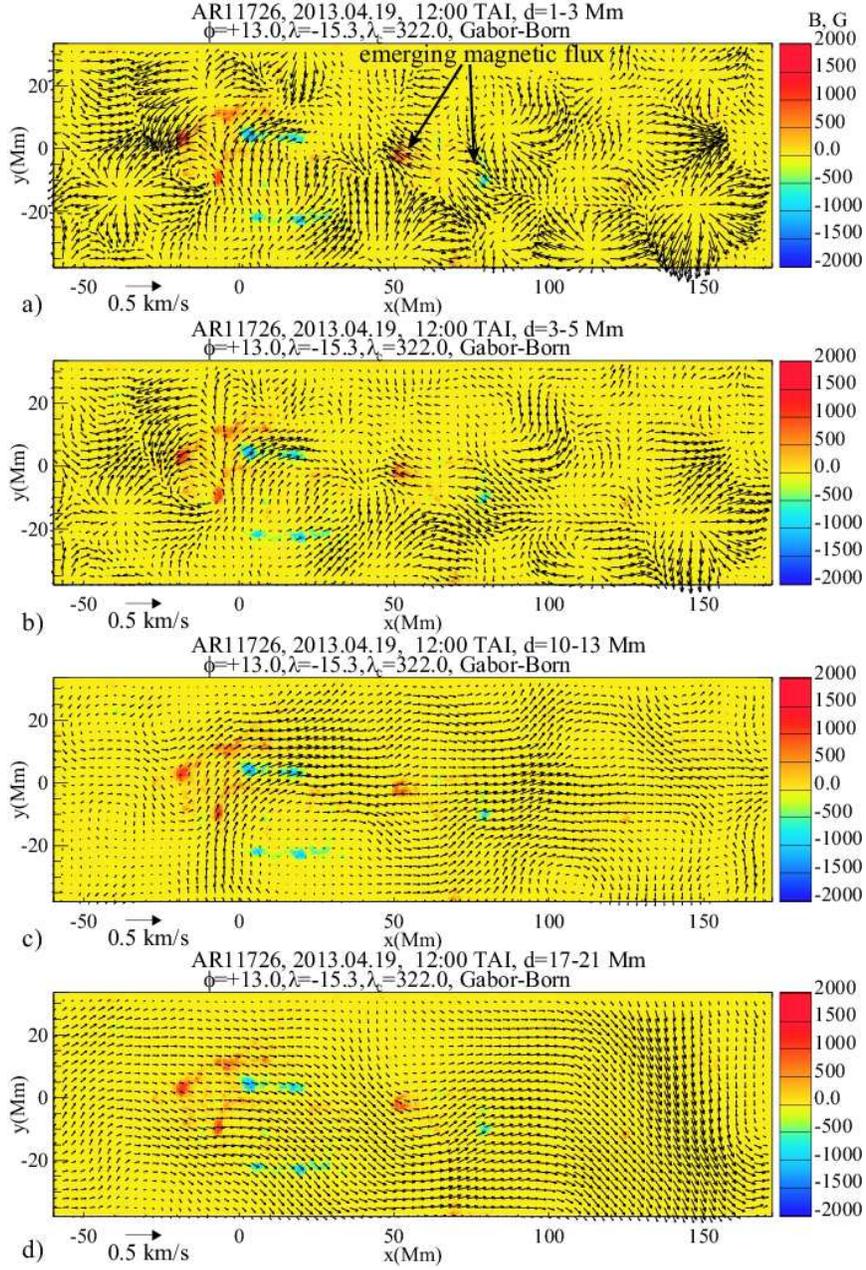}
\caption{Subsurface flow maps and the photospheric magnetogram at the initial moment of appearance of emerging bipolar magnetic flux on the solar surface at 12:00~UT, April 19, 2013, about 12 hours after the initial detection of the subsurface perturbation at four depth: a) $1-3$~Mm, b) $3-5$~Mm, c) $10-13$~Mm, and d) $17-21$~Mm.  
The point $x=0, y=0$ is located at the  heliographic coordinates: latitude $\phi = 13^{\circ}$, longitude $\lambda = -15.3^{\circ}$, and the Carrington longitude $\lambda_c = 322^{\circ}$. The travel times were calculated by using the Gabor-wavelet fitting technique, and the inversion for flows was performed by using the Born-approximation kernels.
}\label{fig3}\end{center}\end{figure}

\begin{figure}
\begin{center}
  \includegraphics[width=0.7\linewidth]{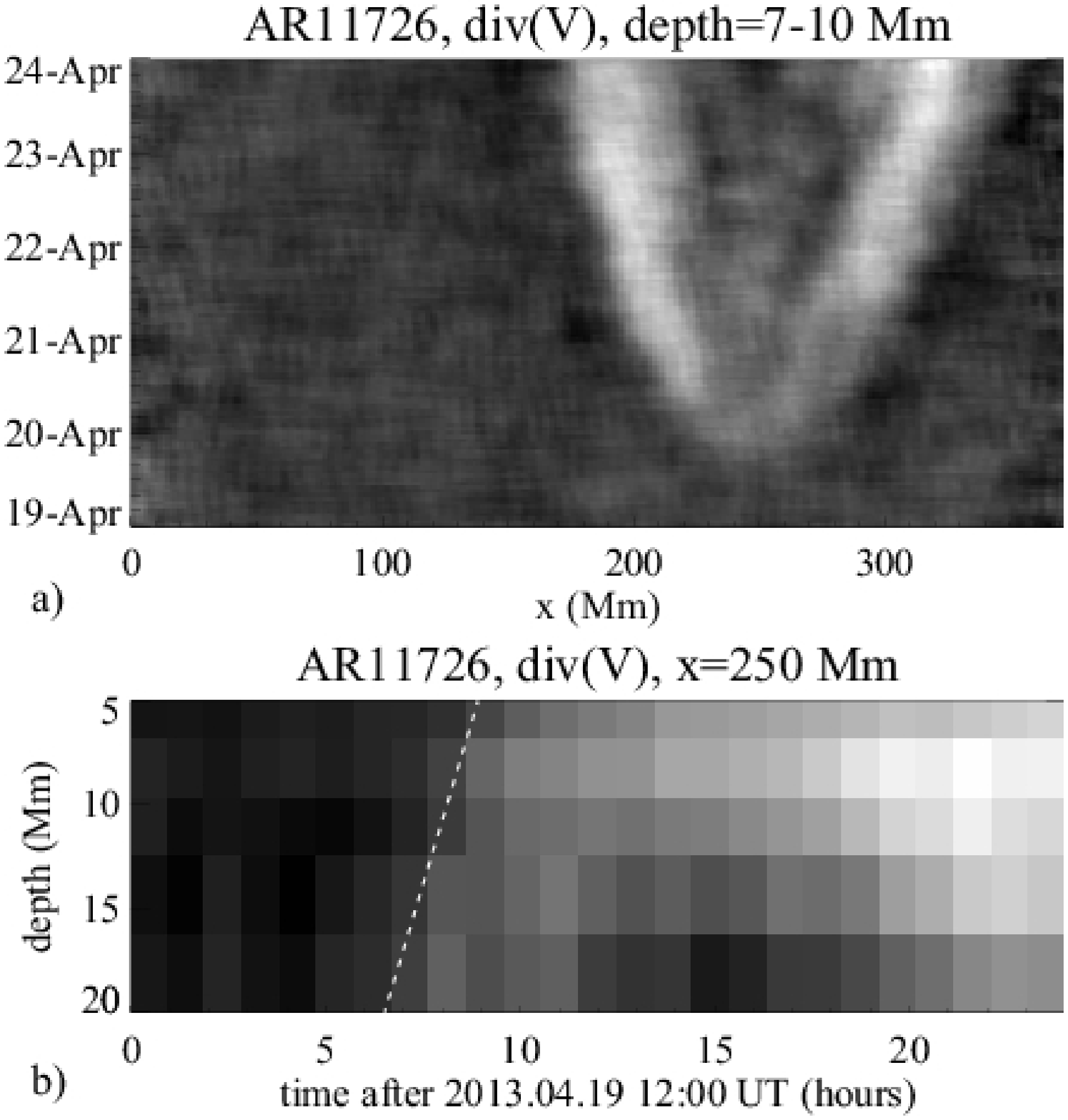}
  \caption{a) Evolution of the horizontal flow divergence in the depth range of $7-10$~Mm is shown as a slice along the longitudinal coordinate $x$  at $y=0$. b) the evolution of the horizontal divergence as a function of depth at the horizontal coordinates: $x=250$~Mm, $y=0$, corresponding the earliest perturbation in panel $a$.   
}\label{fig4}
\end{center}
\end{figure}
Figure~\ref{fig3} showing  the horizontal flow maps at  four different depth at the initial emergence of a bipolar magnetic structure on April 19, 2012, 12:00~UT, overlaid over the photospheric magnetogram. In the near-surface  $1-5$~Mm deep layers (Fig~\ref{fig3}a-b) the  magnetic flux appeared near the boundaries of a supergranulation cell. However, in the deeper layers, at the depth of $10-21$~Mm, no characteristic flow pattern associated with the emerging flux can be visually identified. 

Thus, the near surface flow maps prior the emergence do not reveal a distinct large-scale flow pattern which would indicate that a large magnetic structure is coming up.  Nevertheless these maps allow us to track the process of emergence. Figure~\ref{fig4}a shows a time-space slice of divergence of the horizontal velocity through the subsurface layers in the East-West direction in the depth interval of $7-10$~Mm. The initial perturbation associated with the flux emergence appeared in our domain, centered at Carrington longitude of $322.0$~degrees and $-15.3$ degrees latitude, at $x_0 \approx 250$~Mm. The perturbation represents diverging flows localized around the emerging $\Omega$-loop like structure. In Figure~\ref{fig4}b we plot the divergence at $x_0$ as a function of time and depth. It shows that the emergence speed (indicated by the inclined dotted white line) is about $1.4$~km/s which is very similar to the speed observed in the deep convection zone. 

\begin{figure}
\begin{center}
  \includegraphics[width=\linewidth]{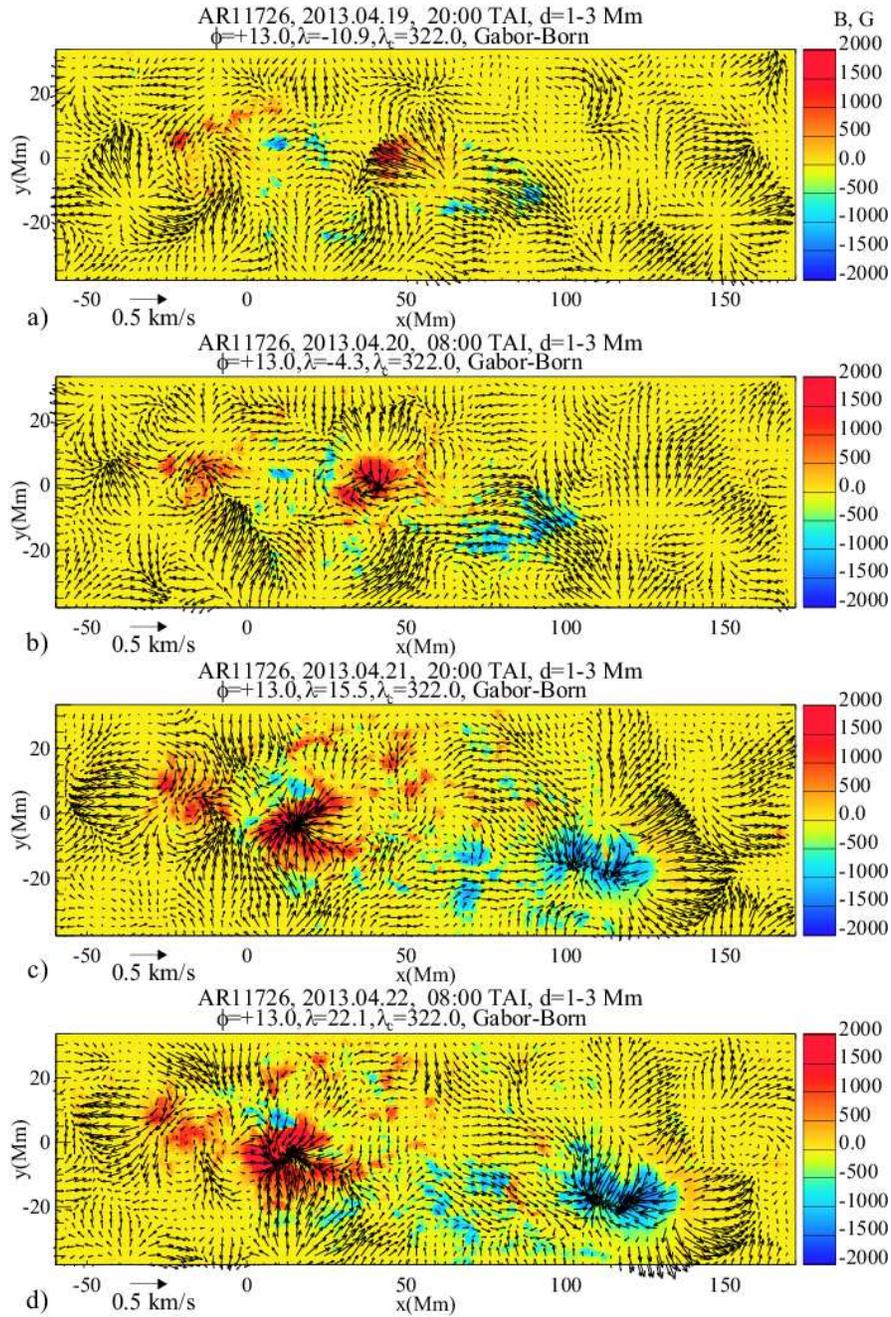}
  \caption{Evolution of  subsurface flows during the emergence of active region NOAA 11726 in the depth range 1-3~Mm for four different moments of time: a)  2013.04.19, 20:00 UT; b) 2013.04.20, 08:00 UT, c) 2013.04.21, 20:00 UT, d) 2013.04.22, 08:00 UT. The corresponding surface magnetograms from HMI are shown in the color background.}\label{fig5}
\end{center}
\end{figure}
\begin{figure}
\begin{center}
  \includegraphics[width=\linewidth]{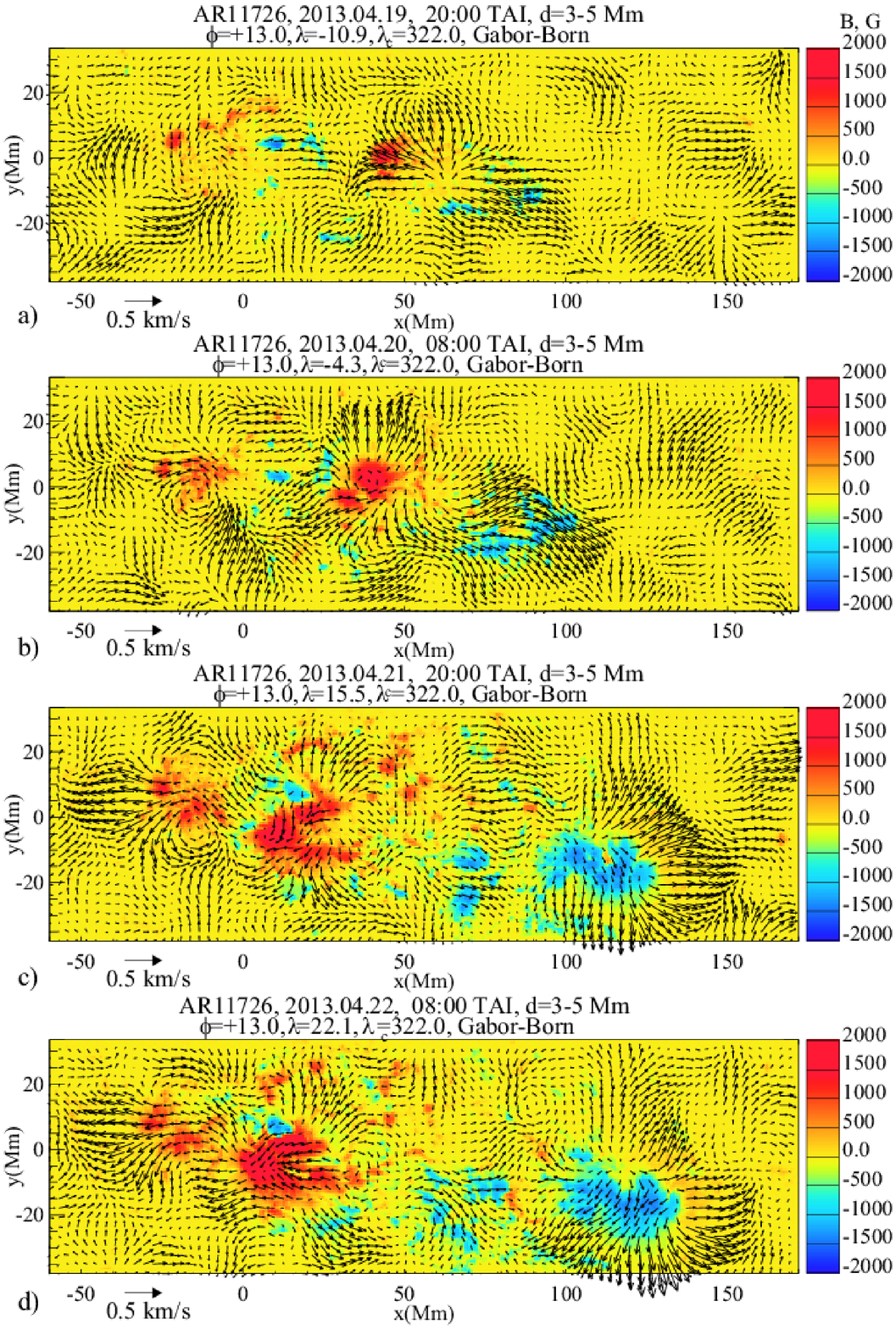}
  \caption{The same as in Fig.~\ref{fig5} for the depth of 3-5~Mm. }\label{fig6}
\end{center}
\end{figure}
\begin{figure}
\begin{center}
  \includegraphics[width=\linewidth]{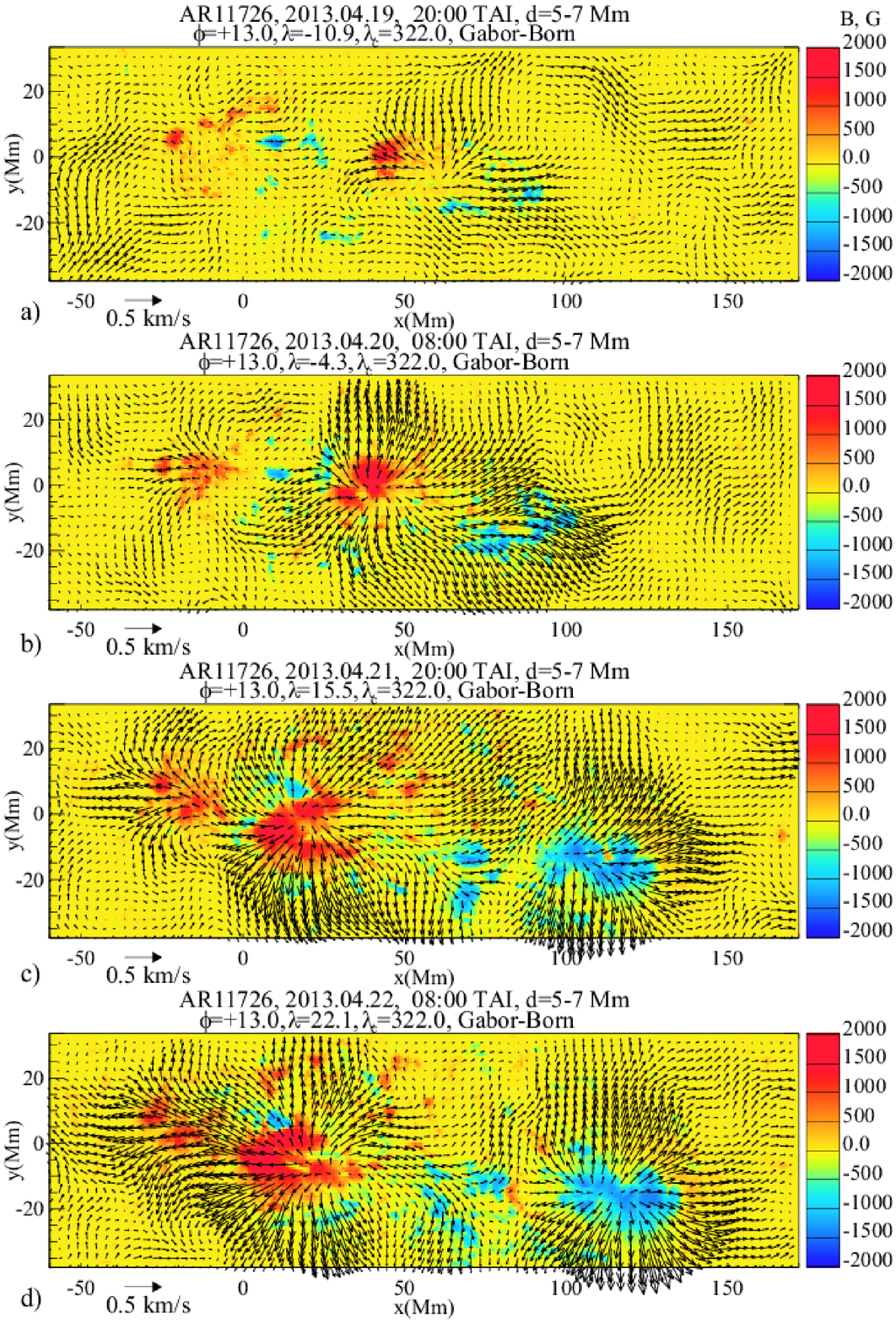}
  \caption{The same as in Fig.~\ref{fig5} for the depth of 5-7~Mm. }\label{fig7}
\end{center}
\end{figure}
\begin{figure}
\begin{center}
  \includegraphics[width=\linewidth]{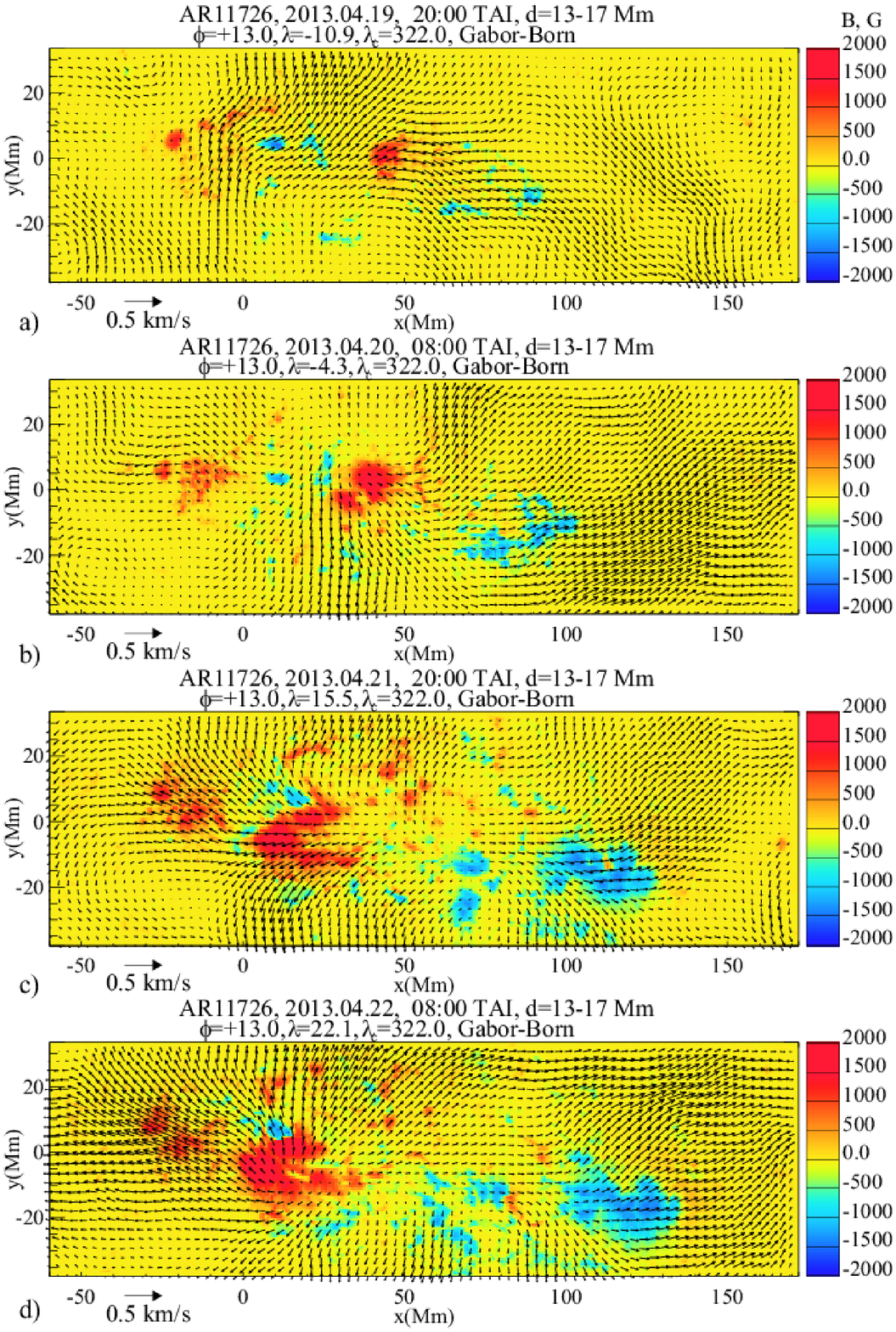}
  \caption{ The same as in Fig.~\ref{fig6} for the depth of $13-17$~Mm. }\label{fig8}
\end{center}
\end{figure}

During the further process the subsurface flows are substantially affected by the  emerging magnetic field. The flow evolution in four different depth ranges is illustrated in Figures~\ref{fig5}-\ref{fig8}. In the subsurface layer, $1-3$~Mm deep,  the initial flow pattern corresponds to the two flux concentrations moving away from each other   (Fig.~\ref{fig5}a). However, 12 hours later we observe formation of converging flows around the positive polarity, associated with the formation of a sunspot (Fig.~\ref{fig5}b). In later times, a similar converging flow pattern is established beneath the leading negative polarity, and is also associated with the formation of sunspots (Fig.~\ref{fig5}c-d). Such converging flows have been observed in the time-distance analysis of the SOHO/MDI Doppler velocity and Hinode/SOT intensity data by using the ray-approximation kernels and a different inversion technique  \citep{Zhao2001,Zhao2010,Zhao2003}. The converging flow pattern beneath the sunspots is very stable and supports the cluster model of sunspots suggested by \citet{Parker1979}. Around the sunspots, the flows become diverging, and are probably associated with the horizontal expansion of the active region. From the evolution of the photospheric magnetic field it is clear that the leading polarity is pushed forward, and the following polarity is moved backward. One can also notice that the between the polarities, in the middle of the active region, the horizontal flows are suppressed.

In the deeper layer  ($3-5$~Mm), the flows are mostly diverging and are concentrated at the boundaries of the active region. Beneath the following sunspot the flows are weaker  but still converging (Fig.~\ref{fig6}). However, beneath the leading polarity the diverging flows dominate. At the depth of $5-7$~Mm the diverging  flow pattern is dominant around both, the leading and following sunspot (Fig.~\ref{fig7}). While in this range of depth the diverging flows are localized around the individual magnetic structures, at greater depths the diverging flow surrounds the whole active region and extends to larger distances, as illustrated in Fig.~\ref{fig8} that shows the flows in the $13-17$~Mm deep layer. Perhaps,  this corresponds to the increasing horizontal extend of the subsurface magnetic region with depth.

\section{Formation of large-scale inflows}\label{sec5}

Previous investigations by the ring-diagram technique \citep{Haber2003} and time-distance helioseismology \citep{Kosovichev1996} revealed large-scale converging flows around active regions, which alter the mean meridional circulation \citep{Haber2002,Zhao2004,Kosovichev2016} and, thus, the magnetic flux transport affecting the strength and duration of the solar activity cycles. The detailed flow maps from the HMI Time-Distance Helioseismology Pipeline allow us to investigate the process of formation of these flows during the emergence of active regions.

\begin{figure}
\begin{center}
  \includegraphics[width=\linewidth]{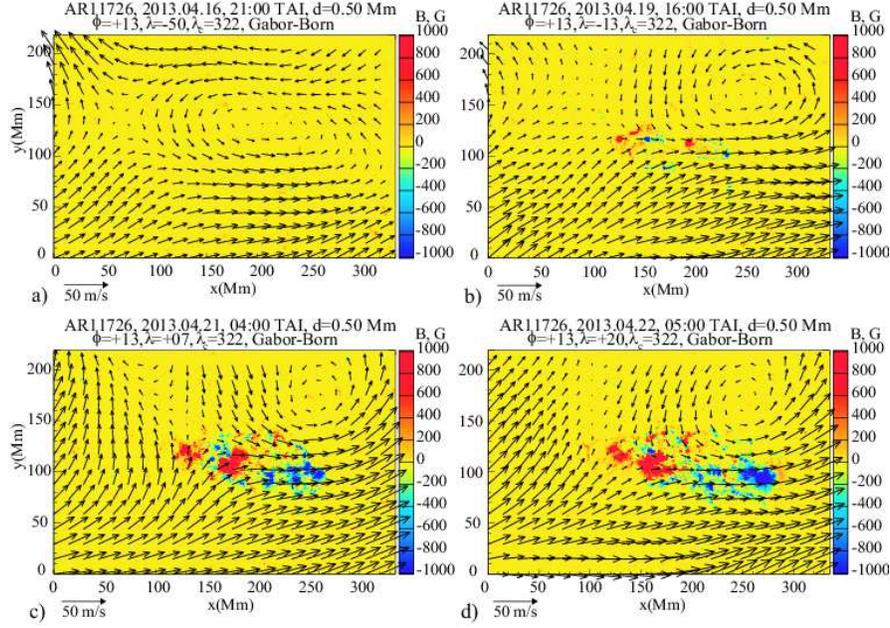}
  \caption{Formation of the large-scale converging flow pattern during the active emergence illustrated for four moments of time: a) 2013.04.16 21:00 UT; b) 2013.04.19 16:00 UT; c) 2013.04.21 04:00 UT; d) 2013.04.22 05:00 UT. The flow maps are obtained by applying a Gaussian smoothing filter with the standard deviation of 30~Mm.    }\label{fig9}
\end{center}
\end{figure}


To isolate large-scale flow patterns from the full-resolution flow maps, samples of which are shown in Figures~\ref{fig5}-\ref{fig8}, we applied a Gaussian smoothing filter with the standard deviation of 30~Mm. This filter smooths the supergranulation-size flows and reveals larger-scale patterns. The result of this filtering applied to the area of emergence of AR~11726 are shown in  Figure~\ref{fig9}. Prior the emergence the subsurface flow pattern represents a shearing zonal flow (Fig.~\ref{fig9}a). At the start of the emergence the zonal flows on the both sides of the emerging magnetic flux are diverted towards the active region, forming a vortex-like structure in the northern part of the area (Fig.~\ref{fig9}b). Then, the converging inflows are amplified and remain stable when most of the flux had emerged on the surface (Fig.~\ref{fig9}c-d).  
\begin{figure}
\begin{center}
  \includegraphics[width=0.7\linewidth]{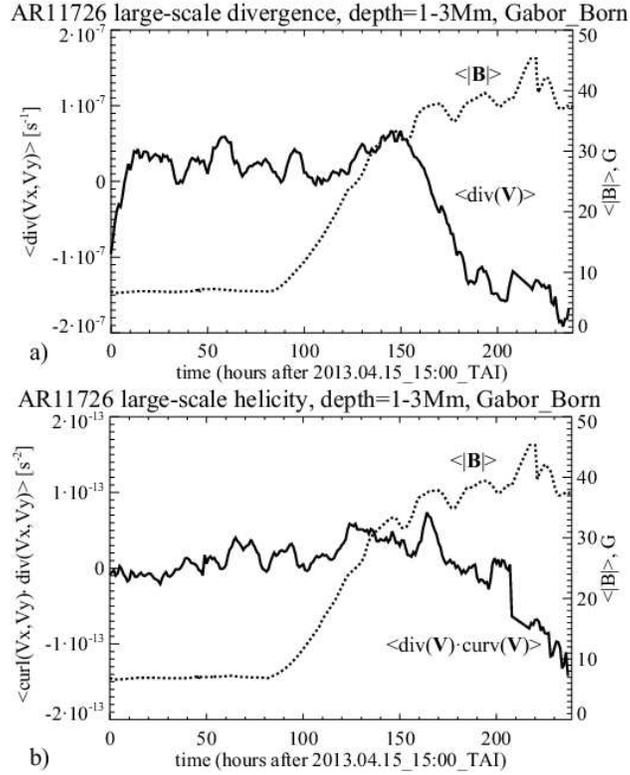}
  \caption{Evolution of a) the mean flow divergence, and b) the mean kinetic helicity beneath the emerging active region NOAA 11726  at the depth of 1-3~Mm.   }\label{fig10}
\end{center}
\end{figure}

To investigate the relationship between the flux emergence and formation of the converging inflow in Figure~\ref{fig10}a we compare the mean flow divergence with the mean unsigned magnetic field strength, calculated for the area shown in Fig.~\ref{fig9}. It shows that the mean divergence became sharply negative after most of the magnetic flux emerged on the surface. The process of formation of the converging flow took about 24 hours. Figure~\ref{fig10}b shows a similar comparison for the kinetic helicity proxy: $\left< (\nabla\cdot\vecV)\cdot(\nabla\times\vecV)_z\right>$. However, while the divergence value quickly saturates the helicity value keeps increasing. The helicity increase means that a large-scale vortex structure is formed beneath the active region. 

The previous study of \citet{Kosovichev2016} showed that the large-scale inflows affect the mean meridional flow in a 10~Mm-deep layer at the top of the convection zone, by effectively reducing the flow speed at 20-40~degrees latitude. As the solar cycle progresses the zone of the reduced speed migrates towards the equator. However, the mean helicity proxy has a strong hemispheric asymmetry (being positive in the Northern hemisphere), and remains largely unchanged during the solar cycle. This, probably means that most of the helicity value comes from the supergranulation. The helicity associated with the active region flows has the opposite sign to the mean helicity but contributes only a few percent. 

These initial results show that the subsurface flows that develop in and around of emerging active regions have a complex multi-scale structure  which is important  for the understanding of how the active regions are formed and how they affect the global Sun's dynamics and magnetic activity. 

\section{Synoptic flow maps}\label{sec6}

To get some insight in the global structure of large-scale flows on the Sun, sometimes called `Solar Subsurface Weather'  \citep{Haber2003} we used the smoothed large-scale flow maps to construct synoptic flow maps similar to the Carrington magnetic field maps. For this, we applied a $\cos^4(1.5\phi)$ filter centered at the central meridian with latitude $\phi=0$ to the individual full-disk flow maps smoothed with the Gaussian filter as described in Section~\ref{sec5}. Then, the filtered flow maps calculated with the 8-hour cadence were combined into the Carrington rotation maps by assigning the appropriate Carrington longitude to the central meridian of the individual flow maps.

\begin{figure}
\begin{center}
  \includegraphics[width=\linewidth]{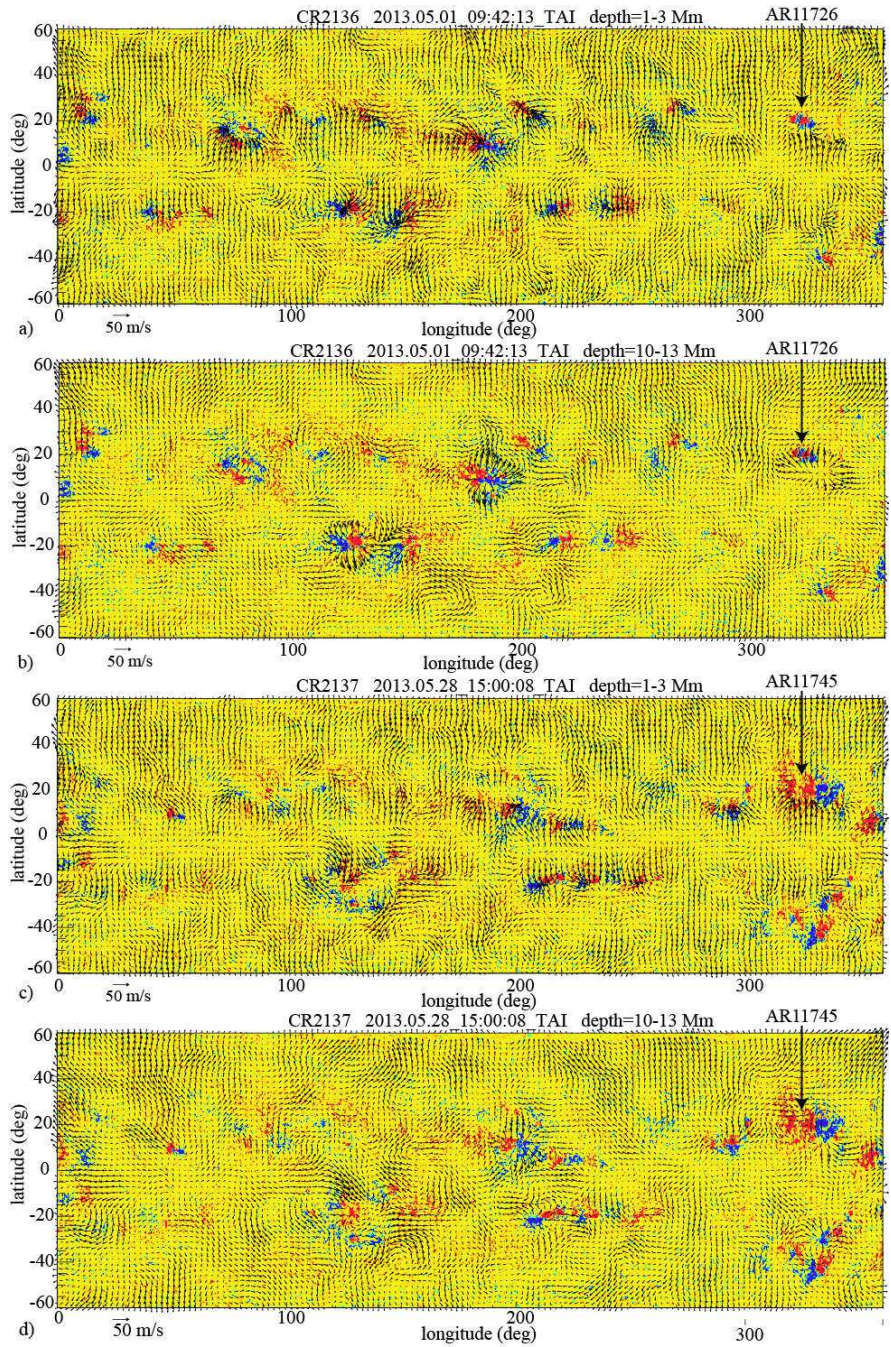}
  \caption{Subsurface synoptic flow maps for two consecutive Carrington rotations at two different depth intervals: a) CR 2166 at $1-3$~Mm, b) CR 2166 at $10-13$~Mm; c) CR 2167 at $1-3$~Mm; and d) CR 2167 at $10-13$~Mm. The corresponding synoptic maps for the radial magnetic field component are shown in the background. Location of the emerging active region NOAA~11726 is indicated by arrow.}\label{fig11}
\end{center}
\end{figure}
Figure~\ref{fig11}a shows the synoptic flow map for Carrington Rotation 2136 and the depth range of $1-3$~Mm. The color background in the corresponding HMI synoptic map for the radial magnetic field. This Carrington Rotation includes the AR~11726 presented in the previous sections (indicated by arrow in Fig.~\ref{fig11}). During its passage through the central meridian (approximately at 16:00~UT, 2013.04.20) the active region was in the middle of the flux emergence process when the large-scale converging flows were still forming. The converging flows are clearly seen around other active regions on this map. In the deeper layers ($10-13$~Mm) shown in Fig.~\ref{fig11}b we observe very prominent outflows two active regions. However, the outflow associated with AR~11726 is not centered on this region and displaced to the south of it. Perhaps, this is a signature of the continuing flux emergence. Indeed, the synoptic maps of the following Carrington Rotation 2137 shows AR~11726 grew into a very large active region which has a new NOAA number 11745  (Fig.~\ref{fig11}c-d).   Of course, a substantial statistical study is required to determine whether the deep subsurface flow patterns can be used for prediction of further evolution of active regions. 

The synoptic flow maps reveal also persistent flow patterns on the scale substantially larger than the scale of the active region flows. An apparent correlation of the flows with the distribution of large-scale fields outside active regions is particularly intriguing, and requires further investigation. 

\section{Conclusion}

As a case study we presented a local helioseismology analysis of the subsurface dynamics of emerging active region NOAA~11726 which is the largest emerging region observed by the SDO/HMI instrument during the first five years of operation. The active region emergence was detected at the depth of $62-75$~Mm about 12 hours  before the first bipolar magnetic structure appeared on the surface, and 2 days before the emergence of most of the magnetic flux. The characteristic speed of emergence estimated from the signal delay in two layers, $62-75$~Mm and $42-55$~Mm deep, is about 1.4~km/s. 

During emergence of the initial bipolar structure no specific large-scale flow pattern in the depth range $0-20$~Mm were identified. Nevertheless, the region of the flux emergence is characterized by an enhanced horizontal flow divergence that corresponds to the spatial separation of the magnetic polarities. The speed of emergence determined by tracking the initial divergence signal with depth is about $1.4$~km/s, very close to the emergence speed in the deep layers.  

As the emerging magnetic flux becomes concentrated in sunspots local converging flows are observed beneath the forming sunspots. The converging flows are most prominent in the depth range $1-3$~Mm, and remain converging after the formation process is completed. The structure of the converging flows is complicated and apparently reflects the sunspot structural evolution and interaction with the surrounding convection flows. The characteristic speed of these flows is about 0.3~km/s. In the deeper layers the flows beneath  the sunspots are predominantly diverging and occupy larger areas. At the depth about 15~Mm the diverging flows occupy a large area around the whole active region. 

By applying a Gaussian filter to smooth the supergranulation-scale flows we investigated the formation of large-scale converging flows around the active region. The scale of these flows is much larger than the size of the active region, and the typical flow speed is about 30~m/s. The formation of the converging flows appears as a diversion of the zonal shearing flows towards the active region, accompanied by formation of a large-scale vortex structure. This process occurs when a substantial amount of the magnetic flux emerged on the surface, and the converging flow pattern remains stable during the following evolution of the active region. The flow helicity is opposity in sign to the subsurface hemispheric helicity mostly determined by the supergranulation flows, but does not significantly contributes to the helicity balance. The primary effect of the converging large-scale flows is in changing the speed of the mean meridional flow in the top 10~Mm-deep layer. 

The synoptic flow maps presented for the Carrington rotation that includes our case study of AR~11726 show that the large-scale subsurface inflows are typical for other active regions. In the deeper layers ($10-13$~Mm) the flows become diverging, and quite strong beneath some active regions. The active region 11726 in the synoptic map is presented at the beginning of the emergence, but  in the deep layers is accompanied by an area of strong divergence, which is off-side of the emerged flux. It remains to be seen if the deep diverging flows indicate on the future development of active regions, but the synoptic map of the following rotation shows that the active region continued to grow on the far side of the Sun and became very large (it received the new NOAA number 11745). 

In addition to the flows around active regions the synoptic maps reveal a complex evolving pattern of large-scale flows on the scale much larger than supergranulation. It appears that these flows correlate with the large-scale magnetic field outside active region. The exact relationship has not been established, but the presented case study encourages further in-depth investigations of the solar subsurface dynamics, both observationally and by numerical MHD simulations.

\section*{Acknowledgment}
This work was supported by the NASA grant NNX14AB70G .

\addcontentsline{toc}{section}{References}
\bibliographystyle{spbasic}

\begin{thebibliography}{39}
\providecommand{\natexlab}[1]{#1}
\providecommand{\url}[1]{{#1}}
\providecommand{\urlprefix}{URL }
\expandafter\ifx\csname urlstyle\endcsname\relax
  \providecommand{\doi}[1]{DOI~\discretionary{}{}{}#1}\else
  \providecommand{\doi}{DOI~\discretionary{}{}{}\begingroup
  \urlstyle{rm}\Url}\fi
\providecommand{\eprint}[2][]{\url{#2}}

\bibitem[{{Birch} and {Gizon}(2007)}]{Birch2007}
{Birch} AC, {Gizon} L (2007) Linear sensitivity of helioseismic travel times to
  local flows. Astronomische Nachrichten 328:228, \doi{10.1002/asna.200610724},
  \urlprefix\url{http://adsabs.harvard.edu/abs/2007AN....328..228B},
  \eprint{1002.2338}

\bibitem[{{Birch} and {Kosovichev}(2000)}]{Birch2000}
{Birch} AC, {Kosovichev} AG (2000) Travel time sensitivity kernels. \solphys
  192:193--201, \doi{10.1023/A:1005283526062},
  \urlprefix\url{http://adsabs.harvard.edu/abs/2000SoPh..192..193B}

\bibitem[{{Birch} and {Kosovichev}(2001)}]{Birch2001a}
{Birch} AC, {Kosovichev} AG (2001) The born approximation in time-distance
  helioseismology. In: {Wilson} A, {Pall{\'e}} PL (eds) SOHO 10/GONG 2000
  Workshop: Helio- and Asteroseismology at the Dawn of the Millennium, ESA
  Special Publication, vol 464, pp 187--190,
  \urlprefix\url{http://adsabs.harvard.edu/abs/2001ESASP.464..187B}

\bibitem[{{Birch} et~al(2001){Birch}, {Kosovichev}, {Price}, and
  {Schlottmann}}]{Birch2001}
{Birch} AC, {Kosovichev} AG, {Price} GH, {Schlottmann} RB (2001) The accuracy
  of the born and ray approximations in time-distance helioseismology. \apjl
  561:L229--L232, \doi{10.1086/324766},
  \urlprefix\url{http://adsabs.harvard.edu/abs/2001ApJ...561L.229B}

\bibitem[{{Birch} et~al(2004){Birch}, {Kosovichev}, and {Duvall}}]{Birch2004}
{Birch} AC, {Kosovichev} AG, {Duvall} TL Jr (2004) Sensitivity of acoustic wave
  travel times to sound-speed perturbations in the solar interior. \apj
  608:580--600, \doi{10.1086/386361},
  \urlprefix\url{http://adsabs.harvard.edu/abs/2004ApJ...608..580B}

\bibitem[{{Birch} et~al(2013){Birch}, {Braun}, {Leka}, {Barnes}, and
  {Javornik}}]{Birch2013}
{Birch} AC, {Braun} DC, {Leka} KD, {Barnes} G, {Javornik} B (2013)
  Helioseismology of pre-emerging active regions. ii. average emergence
  properties. \apj 762:131, \doi{10.1088/0004-637X/762/2/131},
  \eprint{1303.1391}

\bibitem[{{Brun} et~al(2004){Brun}, {Miesch}, and {Toomre}}]{Brun2004}
{Brun} AS, {Miesch} MS, {Toomre} J (2004) Global-scale turbulent convection and
  magnetic dynamo action in the solar envelope. \apj 614:1073--1098,
  \doi{10.1086/423835}, \eprint{astro-ph/0610073}

\bibitem[{{Couvidat} et~al(2006){Couvidat}, {Birch}, and
  {Kosovichev}}]{Couvidat2006}
{Couvidat} S, {Birch} AC, {Kosovichev} AG (2006) Three-dimensional inversion of
  sound speed below a sunspot in the born approximation. \apj 640:516--524,
  \doi{10.1086/500103},
  \urlprefix\url{http://adsabs.harvard.edu/abs/2006ApJ...640..516C}

\bibitem[{{Couvidat} et~al(2012){Couvidat}, {Zhao}, {Birch}, {Kosovichev},
  {Duvall}, {Parchevsky}, and {Scherrer}}]{Couvidat2012}
{Couvidat} S, {Zhao} J, {Birch} AC, {Kosovichev} AG, {Duvall} TL, {Parchevsky}
  K, {Scherrer} PH (2012) Implementation and comparison of acoustic travel-time
  measurement procedures for the solar dynamics observatory/helioseismic and
  magnetic imager time - distance helioseismology pipeline. \solphys
  275:357--374, \doi{10.1007/s11207-010-9652-y},
  \urlprefix\url{http://adsabs.harvard.edu/abs/2012SoPh..275..357C}

\bibitem[{{D'Silva} and {Choudhuri}(1993)}]{DSilva1993}
{D'Silva} S, {Choudhuri} AR (1993) A theoretical model for tilts of bipolar
  magnetic regions. \aap 272:621

\bibitem[{{Fan} and {Fang}(2014)}]{Fan2014}
{Fan} Y, {Fang} F (2014) A simulation of convective dynamo in the solar
  convective envelope: Maintenance of the solar-like differential rotation and
  emerging flux. \apj 789:35, \doi{10.1088/0004-637X/789/1/35},
  \eprint{1405.3926}

\bibitem[{{Gizon} and {Birch}(2002)}]{Gizon2002}
{Gizon} L, {Birch} AC (2002) Time-distance helioseismology: The forward problem
  for random distributed sources. \apj 571:966--986, \doi{10.1086/340015},
  \urlprefix\url{http://adsabs.harvard.edu/abs/2002ApJ...571..966G}

\bibitem[{{Guerrero} et~al(2016){Guerrero}, {Smolarkiewicz}, {de Gouveia Dal
  Pino}, {Kosovichev}, and {Mansour}}]{Guerrero2016}
{Guerrero} G, {Smolarkiewicz} PK, {de Gouveia Dal Pino} EM, {Kosovichev} AG,
  {Mansour} NN (2016) On the role of tachoclines in solar and stellar dynamos.
  \apj 819:104, \doi{10.3847/0004-637X/819/2/104}, \eprint{1507.04434}

\bibitem[{{Haber} et~al(2002){Haber}, {Hindman}, {Toomre}, {Bogart}, {Larsen},
  and {Hill}}]{Haber2002}
{Haber} DA, {Hindman} BW, {Toomre} J, {Bogart} RS, {Larsen} RM, {Hill} F (2002)
  Evolving submerged meridional circulation cells within the upper convection
  zone revealed by ring-diagram analysis. \apj 570:855--864,
  \doi{10.1086/339631},
  \urlprefix\url{http://adsabs.harvard.edu/abs/2002ApJ...570..855H}

\bibitem[{{Haber} et~al(2003){Haber}, {Hindman}, and {Toomre}}]{Haber2003}
{Haber} DA, {Hindman} BW, {Toomre} J (2003) Interaction of solar subsurface
  flows with major active regions. In: {Sawaya-Lacoste} H (ed) GONG+ 2002.
  Local and Global Helioseismology: the Present and Future, ESA Special
  Publication, vol 517, pp 103--108,
  \urlprefix\url{http://adsabs.harvard.edu/abs/2003ESASP.517..103H}

\bibitem[{{Hindman} et~al(2009){Hindman}, {Haber}, and {Toomre}}]{Hindman2009}
{Hindman} BW, {Haber} DA, {Toomre} J (2009) Subsurface circulations within
  active regions. \apj 698:1749--1760, \doi{10.1088/0004-637X/698/2/1749},
  \eprint{0904.1575}

\bibitem[{{Ilonidis} et~al(2013){Ilonidis}, {Zhao}, and
  {Hartlep}}]{Ilonidis2013}
{Ilonidis} S, {Zhao} J, {Hartlep} T (2013) Helioseismic investigation of
  emerging magnetic flux in the solar convection zone. \apj 777:138,
  \doi{10.1088/0004-637X/777/2/138},
  \urlprefix\url{http://adsabs.harvard.edu/abs/2013ApJ...777..138I}

\bibitem[{{Jacobsen} et~al(1999){Jacobsen}, {Moller}, {Jensen}, and
  {Efferso}}]{Jacobsen1999}
{Jacobsen} B, {Moller} I, {Jensen} J, {Efferso} F (1999) Multichannel
  deconvolution, mcd, in geophysics and helioseismology. Physics and Chemistry
  of the Earth A 24:215--220, \doi{10.1016/S1464-1895(99)00021-6},
  \urlprefix\url{http://adsabs.harvard.edu/abs/1999PCEA...24..215J}

\bibitem[{{K{\"a}pyl{\"a}} et~al(2016){K{\"a}pyl{\"a}}, {Brandenburg},
  {Kleeorin}, {K{\"a}pyl{\"a}}, and {Rogachevskii}}]{Kapyla2016}
{K{\"a}pyl{\"a}} PJ, {Brandenburg} A, {Kleeorin} N, {K{\"a}pyl{\"a}} MJ,
  {Rogachevskii} I (2016) Magnetic flux concentrations from turbulent
  stratified convection. \aap 588:A150, \doi{10.1051/0004-6361/201527731},
  \eprint{1511.03718}

\bibitem[{{Kitiashvili} et~al(2010){Kitiashvili}, {Kosovichev}, {Wray}, and
  {Mansour}}]{Kitiashvili2010}
{Kitiashvili} IN, {Kosovichev} AG, {Wray} AA, {Mansour} NN (2010) Mechanism of
  spontaneous formation of stable magnetic structures on the sun. \apj
  719:307--312, \doi{10.1088/0004-637X/719/1/307}, \eprint{1004.2288}

\bibitem[{{Komm} et~al(2011){Komm}, {Howe}, and {Hill}}]{Komm2011}
{Komm} R, {Howe} R, {Hill} F (2011) Subsurface velocity of emerging and
  decaying active regions. \solphys 268:407--428,
  \doi{10.1007/s11207-010-9692-3}

\bibitem[{{Komm} et~al(2012){Komm}, {Howe}, and {Hill}}]{Komm2012}
{Komm} R, {Howe} R, {Hill} F (2012) Vorticity of subsurface flows of emerging
  and decaying active regions. \solphys 277:205--226,
  \doi{10.1007/s11207-011-9920-5}

\bibitem[{{Kosovichev}(1996)}]{Kosovichev1996}
{Kosovichev} AG (1996) Tomographic imaging of the sun's interior. \apjl
  461:L55, \doi{10.1086/309989},
  \urlprefix\url{http://adsabs.harvard.edu/abs/1996ApJ...461L..55K}

\bibitem[{{Kosovichev}(2012)}]{Kosovichev2012}
{Kosovichev} AG (2012) Local helioseismology of sunspots: Current status and
  perspectives. \solphys 279:323--348, \doi{10.1007/s11207-012-9996-6},
  \urlprefix\url{http://adsabs.harvard.edu/abs/2012SoPh..279..323K}

\bibitem[{{Kosovichev} and {Duvall}(2006)}]{Kosovichev2006}
{Kosovichev} AG, {Duvall} TL (2006) Active region dynamics. \ssr 124:1--12,
  \doi{10.1007/s11214-006-9112-z},
  \urlprefix\url{http://adsabs.harvard.edu/abs/2006SSRv..124....1K}

\bibitem[{{Kosovichev} and {Duvall}(1997)}]{Kosovichev1997a}
{Kosovichev} AG, {Duvall} TL Jr (1997) Acoustic tomography of solar convective
  flows and structures. In: {Pijpers} FP, {Christensen-Dalsgaard} J,
  {Rosenthal} CS (eds) SCORe'96 : Solar Convection and Oscillations and their
  Relationship, Astrophysics and Space Science Library, vol 225, pp 241--260,
  \urlprefix\url{http://adsabs.harvard.edu/abs/1997ASSL..225..241K}

\bibitem[{{Kosovichev} and {Zhao}(2016)}]{Kosovichev2016}
{Kosovichev} AG, {Zhao} J (2016) Reconstruction of solar subsurfaces by local
  helioseismology. In: {Rozelot} JP, {Neiner} C (eds) Lecture Notes in
  Physics,Springer International Publishing Switzerland, Lecture Notes in
  Physics, Springer International Publishing, Switzerland, vol 914, pp 25--41

\bibitem[{{Masada} and {Sano}(2016)}]{Masada2016}
{Masada} Y, {Sano} T (2016) Spontaneous formation of surface magnetic structure
  from large-scale dynamo in strongly stratified convection. \apjl 822:L22,
  \doi{10.3847/2041-8205/822/2/L22}, \eprint{1604.05374}

\bibitem[{{Parker}(1979)}]{Parker1979}
{Parker} EN (1979) Sunspots and the physics of magnetic flux tubes. i - the
  general nature of the sunspot. ii - aerodynamic drag. \apj 230:905--923,
  \doi{10.1086/157150}

\bibitem[{{Scherrer} et~al(2012){Scherrer}, {Schou}, {Bush}, {Kosovichev},
  {Bogart}, {Hoeksema}, {Liu}, {Duvall}, {Zhao}, {Title}, {Schrijver},
  {Tarbell}, and {Tomczyk}}]{Scherrer2012}
{Scherrer} PH, {Schou} J, {Bush} RI, {Kosovichev} AG, {Bogart} RS, {Hoeksema}
  JT, {Liu} Y, {Duvall} TL, {Zhao} J, {Title} AM, {Schrijver} CJ, {Tarbell} TD,
  {Tomczyk} S (2012) The helioseismic and magnetic imager (hmi) investigation
  for the solar dynamics observatory (sdo). \solphys 275:207--227,
  \doi{10.1007/s11207-011-9834-2},
  \urlprefix\url{http://adsabs.harvard.edu/abs/2012SoPh..275..207S}

\bibitem[{{Woodard}(1984)}]{Woodard1984}
{Woodard} MF (1984) Short-period oscillations in the total solar irradiance.
  PhD thesis, UNIVERSITY OF CALIFORNIA, SAN DIEGO.

\bibitem[{{Zhao} and {Kosovichev}(2003)}]{Zhao2003}
{Zhao} J, {Kosovichev} AG (2003) Helioseismic observation of the structure and
  dynamics of a rotating sunspot beneath the solar surface. \apj 591:446--453,
  \doi{10.1086/375343},
  \urlprefix\url{http://adsabs.harvard.edu/abs/2003ApJ...591..446Z}

\bibitem[{{Zhao} and {Kosovichev}(2004)}]{Zhao2004}
{Zhao} J, {Kosovichev} AG (2004) Torsional oscillation, meridional flows, and
  vorticity inferred in the upper convection zone of the sun by time-distance
  helioseismology. \apj 603:776--784, \doi{10.1086/381489},
  \urlprefix\url{http://adsabs.harvard.edu/abs/2004ApJ...603..776Z}

\bibitem[{{Zhao} et~al(2001){Zhao}, {Kosovichev}, and {Duvall}}]{Zhao2001}
{Zhao} J, {Kosovichev} AG, {Duvall} TL Jr (2001) Investigation of mass flows
  beneath a sunspot by time-distance helioseismology. \apj 557:384--388,
  \doi{10.1086/321491},
  \urlprefix\url{http://adsabs.harvard.edu/abs/2001ApJ...557..384Z}

\bibitem[{{Zhao} et~al(2009){Zhao}, {Kosovichev}, and {Sekii}}]{Zhao2009b}
{Zhao} J, {Kosovichev} AG, {Sekii} T (2009) Subsurface structures and flow
  fields of an active region observed by hinode. In: {Lites} B, {Cheung} M,
  {Magara} T, {Mariska} J, {Reeves} K (eds) The Second Hinode Science Meeting:
  Beyond Discovery-Toward Understanding, Astronomical Society of the Pacific
  Conference Series, vol 415, p 411,
  \urlprefix\url{http://adsabs.harvard.edu/abs/2009ASPC..415..411Z}

\bibitem[{{Zhao} et~al(2010){Zhao}, {Kosovichev}, and {Sekii}}]{Zhao2010}
{Zhao} J, {Kosovichev} AG, {Sekii} T (2010) High-resolution helioseismic
  imaging of subsurface structures and flows of a solar active region observed
  by hinode. \apj 708:304--313, \doi{10.1088/0004-637X/708/1/304},
  \urlprefix\url{http://adsabs.harvard.edu/abs/2010ApJ...708..304Z},
  \eprint{0911.1161}

\bibitem[{{Zhao} et~al(2012){Zhao}, {Couvidat}, {Bogart}, {Parchevsky},
  {Birch}, {Duvall}, {Beck}, {Kosovichev}, and {Scherrer}}]{Zhao2012}
{Zhao} J, {Couvidat} S, {Bogart} RS, {Parchevsky} KV, {Birch} AC, {Duvall} TL,
  {Beck} JG, {Kosovichev} AG, {Scherrer} PH (2012) Time-distance
  helioseismology data-analysis pipeline for helioseismic and magnetic imager
  onboard solar dynamics observatory (sdo/hmi) and its initial results.
  \solphys 275:375--390, \doi{10.1007/s11207-011-9757-y},
  \urlprefix\url{http://adsabs.harvard.edu/abs/2012SoPh..275..375Z},
  \eprint{1103.4646}

\bibitem[{{Zhao} et~al(2013){Zhao}, {Bogart}, {Kosovichev}, {Duvall}, and
  {Hartlep}}]{Zhao2013}
{Zhao} J, {Bogart} RS, {Kosovichev} AG, {Duvall} TL Jr, {Hartlep} T (2013)
  Detection of equatorward meridional flow and evidence of double-cell
  meridional circulation inside the sun. \apjl 774:L29,
  \doi{10.1088/2041-8205/774/2/L29},
  \urlprefix\url{http://adsabs.harvard.edu/abs/2013ApJ...774L..29Z},
  \eprint{1307.8422}

\bibitem[{{Zhao} et~al(2014){Zhao}, {Kosovichev}, and {Bogart}}]{Zhao2014}
{Zhao} J, {Kosovichev} AG, {Bogart} RS (2014) Solar meridional flow in the
  shallow interior during the rising phase of cycle 24. \apjl 789:L7,
  \doi{10.1088/2041-8205/789/1/L7},
  \urlprefix\url{http://adsabs.harvard.edu/abs/2014ApJ...789L...7Z},
  \eprint{1406.2735}

\end{thebibliography}

\end{document}